\documentclass[american,american,preprint]{article}
\usepackage[T1]{fontenc}
\usepackage[utf8]{inputenc}
\usepackage{prettyref}
\usepackage{amstext}

\makeatletter
\usepackage{prettyref}
\newrefformat{eq}{eq.\nolinebreak(\ref{#1})}
\newrefformat{sec}{Section\nolinebreak\,\nolinebreak\ref{#1}}
\newrefformat{cha}{Chapter\nolinebreak\,\nolinebreak\ref{#1}}
\newrefformat{tab}{Table\nolinebreak\,\nolinebreak\ref{#1}}
\newrefformat{fig}{Figure\nolinebreak\,\nolinebreak\ref{#1}}
\newrefformat{lem}{Lemma\nolinebreak\,\nolinebreak\ref{#1}}
\newrefformat{thm}{Theorem\nolinebreak\,\nolinebreak\ref{#1}}
\newrefformat{axm}{Axiom\nolinebreak\,\nolinebreak\ref{#1}}
\newrefformat{app}{Appendix\nolinebreak\,\nolinebreak\ref{#1}}
\newrefformat{exe}{Exercise\nolinebreak\,\nolinebreak\ref{#1}}
\newrefformat{cor}{Corollary\nolinebreak\,\nolinebreak\ref{#1}}
\newrefformat{pro}{Property\nolinebreak\,\nolinebreak\ref{#1}}
\newrefformat{def}{Definition\nolinebreak\,\nolinebreak\ref{#1}}
\newrefformat{subsec}{Section\nolinebreak\,\nolinebreak\ref{#1}}
\usepackage{amsmath}
\usepackage{bm}
\newcommand{\br}[1]{\mathbf{#1}} %bold upright roman
\newcommand{\bs}[1]{\bm{\mathsf{#1}}} %bold sanserif
\newcommand{\bg}[1]{\bm{#1}} %bold greek
\newcommand{\fv}[1]{\bs{#1}} %fourvector

\renewcommand{\@cite}[2]{[{#1\if@tempswa , #2\fi}]}
\renewcommand{\@biblabel}[1]{#1.}

\renewcommand{\section}{\@startsection{section}{1}{0pt}%
{-2.0ex plus -1ex minus -.1ex}{1.0ex plus .1ex}%
{\noindent\normalsize\bf \rule{1pt}{0pt}}}

\renewcommand{\subsection}{\@startsection{subsection}{2}{0pt}%
{-2.0ex plus -1ex minus -.1ex}{1.0ex plus.1ex}%
{\noindent\normalsize\bf \rule{1pt}{0pt}}}

\usepackage{txfonts}
\date{}

\makeatother

\usepackage{babel}
\begin{document}
\onecolumn%

\title{Is Electromagnetic Field Momentum Due to the Flow of Field Energy?\thanks{This is a revised version of an article published in \emph{Studies
in the History and Philosophy of Science} (2021) \textbf{88}: 358-366. }}

\author{{\normalsize{}Oliver Davis Johns}\\
{\normalsize{}San Francisco State University, Physics and Astronomy
Department}\\
{\normalsize{}1600 Holloway Avenue, San Francisco, CA 94133, USA}\\
{\normalsize{}Email: ojohns@metacosmos.org}\\
{\normalsize{}Web: http://www.metacosmos.org}}

\maketitle
\begin{center}%
\begin{minipage}[t]{0.7\columnwidth}%
\qquad{}Conservation laws such as the Poynting theorem of electrodynamics
that are based on the divergence of a second-rank four-tensor, are
fundamentally different from conservation laws such as the conservation
of electric charge that are based on the divergence of a four-vector.
This article investigates the consequences of this difference for
understanding the relation between electromagnetic field momentum
and the flow of electromagnetic field energy.

\qquad{}Momentum and energy conservation require electromagnetic
field momentum and energy to be treated as physically real, even in
static fields. This motivates the conjecture that field momentum might
be due to the flow of a relativistic mass density (defined as energy
density divided by the square of the speed of light).

\vspace{3pt}

\qquad{}This article investigates the velocity of such an energy
flow and finds a conflict between two different definitions of it,
both of which originally seem plausible if the flow is to be taken
as real. This investigation is careful to respect the transformation
rules of special relativity throughout.

\vspace{3pt}

\qquad{}The paper demonstrates that the consensus definition of the
flow velocity of electromagnetic energy is inconsistent with the transformation
rules of special relativity, and hence is incorrect. A correct flow
velocity is then derived which is completely consistent with those
transformation rules. \vspace{3pt}

\qquad{}The conclusion is that these conflicting definitions of energy
flow velocity cannot be resolved in a way that is consistent with
special relativity and that also allows electromagnetic field momentum
density to be the result of relativistic mass flow. Though real, field
momentum density cannot be explained as the flow of field energy.

\vspace{3pt}

\qquad{}As a byproduct of the study, it is also shown that there
is a comoving system in which the electromagnetic energy-momentum
tensor is reduced to a simple diagonal form, with two of its diagonal
elements equal to the energy density and the other two diagonal elements
equal to plus and minus a single parameter derived from the electromagnetic
field values, a result that places constraints on possible fluid models
of electromagnetism.%
\end{minipage}

\end{center}\twocolumn

\section{Introduction\label{sec:Intro}}

The Poynting theorem of electrodynamics is based on the divergence
of the second-rank energy-momentum tensor $T^{\alpha\beta}$. Conservation
laws based on the divergence of a second-rank tensor are fundamentally
different from conservation laws, such as the conservation of electric
charge, that are based on the divergence of a four-vector. This article
investigates the consequences of this difference for understanding
the relation between electromagnetic field momentum and the flow of
electromagnetic field energy.

The example of a rotating disk with a magnet at its center and charged
spheres on its perimeter provides a convincing argument that, to preserve
the principle of angular momentum conservation, the field momentum
of even a static electromagnetic field must be considered physically
real.\footnote{\label{fn:Feynman-et-al}Feynman et al \cite{FeynmanLectures}, Section
17-4, Section 27-6, and Figure 17-5. Quantitative matches of field
to mechanical angular momentum are found, for example, in Romer \cite{Romer-AngMom}
and Boos \cite{Boos-AngMom}.} It is also generally assumed that conservation of energy requires
the energy density of the electromagnetic field to be physically real,
even for static fields. This article accepts the reality of the field
momentum and energy, but questions the flow of field energy as the
source of field momentum.

Explanation of electromagnetic field momentum as an energy flow depends
crucially on a \emph{correct} definition for the velocity of that
energy flow. Since special relativity is the invariance theory of
electromagnetism, throughout this paper by \emph{correct} or \emph{valid
}we will mean that a definition or construct is correct or valid only
if it is consistent with special relativity. 

We give two possible definitions for the velocity of electromagnetic
energy flow, which we refer to as Definitions A and B. For conservation
laws such as charge conservation that are derived from the divergence
of a four-vector, these two definitions coincide. But for conservation
laws such as the Poynting theorem based on the divergence of the second-rank
energy-momentum tensor, the two definitions differ and only one can
be correct.

\textbf{Definition A: }In the modern, post-relativity era it has been
the consensus in the literature, at least since the first English
edition of Born and Wolf's Optics,\footnote{\label{fn:LessC} In a discussion of the Poynting theorem in material
media, but with no special attention to Lorentz covariance, Born and
Wolf \cite{BornWolf} Section 14.2, eq.(8) identify $\br V_{\!\text{A}}$
as the \emph{velocity of energy transport} or \emph{ray velocity.}
(The first edition of Born and Wolf's text appeared in 1959.) Section
B.2 of Smith \cite{Smith} echos Born and Wolf but provides no new
derivation. Geppert \cite{Geppert} makes the same identification.
More recently, Sebens \cite{Sebens-Fields,Sebens-mass} relies on
these and other sources to identify $\br V_{\!\text{A}}$ as the electromagnetic
mass flow velocity. Sebens also considers earlier, pre-Einstein studies
by Poincare. The present paper however is focussed on the reconsideration
of the subject forced by special relativity.} that the electromagnetic energy flow velocity is $\br V_{\!\text{A}}=\br S/{\cal E}$
where $\br S$ is the Poynting vector and ${\cal E}$ is the energy
density. We refer to this as Definition A. 

With Definition A, clearing the fraction to $\br S={\cal E}\br V_{\!\text{A}}$
and dividing by $c^{2}$ gives $\br G={\cal M}_{\text{rel}}\br V_{\!\text{A}}$where
$\br G=\br S/c^{2}$ is the momentum density, and ${\cal M}_{\text{rel}}={\cal E}/c^{2}$
is the so-called relativistic mass density. Thus Definition A implies
that the momentum density is due to flow of relativistic mass. 

This article investigates this consensus claim and finds reason to
doubt it. Sections 2 and 3 demonstrate that the coordinate flow velocity
$\br V_{\!\text{A}}$ is not \emph{correct} in the above sense. It
is not consistent with the transformation rules of special relativity.

\textbf{Definition B:} The velocity of the energy (mass) flow at a
given event can also be defined as the velocity of an observer who
measures the Poynting energy flux vector to be zero at that event.
If $\br S$ truly is the flux of energy flow, then an observer comoving
with this flow should observe a zero value of that energy flux. This
is Definition B and its velocity will be denoted as $\br V_{\!\text{B}}$.
An explicit and relativistically correct derivation of $\br V_{\!\text{B}}$
is presented in \prettyref{sec:DetailB}. Its possible caveats are
discussed in \prettyref{sec:CaveatsB} and \prettyref{sec:Exceptional}.

\prettyref{sec:EnergyMomentum} demonstrates that the comoving reference
system used in the derivation of Definition B allows a reduction of
the electromagnetic energy-momentum tensor to a simple, diagonal form,
with two of its diagonal elements equal to the energy density in the
comoving frame and the other two diagonal elements equal to plus and
minus a single parameter derived from the electromagnetic field values.
This reduction of the electromagnetic energy-momentum tensor is shown
to place important constraints on fluid-dynamic models of energy flow
in the electromagnetic field.

\prettyref{sec:Conclusion} concludes that electromagnetic field momentum
density is not due to the flow of an electromagnetic mass density.
Since $\br V_{\!\text{A}}\neq\br V_{\!\text{B}}$, a choice between
Definitions A and B must be made. As noted above, velocity $\br V_{\!\text{A}}$
would show electromagnetic field momentum to result from the flow
of field energy. However $\br V_{\!\text{A}}$ is not relativistically
correct and must be rejected. However, with the correct and relativistically
valid choice $\br V_{\!\text{B}}$, momentum density\emph{ cannot
}be explained as due to the motion of a relativistic mass density
${\cal M}_{\text{rel}}={\cal E}/c^{2}$. Thus the title question of
this paper has a negative answer.

\prettyref{sec:Maxwell} accepts the negative results of the present
study of relativistic mass flow, and speculates about possible ways
forward in the search for a model, if any, of physics that might underly
the Maxwell Equations.

This paper uses Heaviside-Lorentz units. We denote four-vectors as
$\fv K=K^{0}\fv e_{0}+\br K$ where $\fv e_{0}$ is the time unit
vector and the three-vector part is understood to be $\br K=K^{1}\fv e_{1}+K^{2}\fv e_{2}+K^{3}\fv e_{3}$.
In the Einstein summation convention, Greek indices range from $0$
to $3$, Roman indices from $1$ to $3$. The Minkowski metric tensor
is $(\eta_{\alpha\beta})=(\eta^{\alpha\beta})=\text{diag}(-1,+1,+1,+1)$.
Three-vectors are written with bold type $\br K$, and their magnitudes
as $K$. Thus $\left|\br K\right|=K$.

\section{Arguments for Definition A\label{sec:DetailA}}

Definition A for the energy flow velocity is defined by equivalent\footnote{Since ${\cal E}$ is nonzero except when $E{=}B{=}0$, throughout
this paper we will take Definition A, $\br V_{\!\text{A}}=\br S/{\cal E}$,
to be equivalent to the formula $\br S={\cal E}\br V_{\!\text{A}}$.} formulas 
\begin{equation}
\br S={\cal E}\br V_{\!\text{A}}\quad\quad\text{and}\quad\quad\br V_{\!\text{A}}=\br S/{\cal E}\label{eq:defA3}
\end{equation}

Writing $\br S{=}c\left(\br E\times\br B\right)$ and  $\,\,{\cal E}{=}\left(E^{2}+B^{2}\right)/2$,
in terms of the electric and magnetic fields $\br E$ and $\br B$,
gives\footnote{Electromagnetic formulas in this paper are taken from Griffiths \cite{Griffiths}
and Jackson \cite{Jackson}, with translation into Heaviside-Lorentz
units.}
\begin{equation}
\br V_{\!\text{A}}=\dfrac{\br S}{{\cal E}}=\dfrac{2c\left(\br E\times\br B\right)}{\left(E^{2}+B^{2}\right)}\label{eq:defA1}
\end{equation}
It is easily shown from the inequalities $\left(E-B\right)^{2}\ge0$
and $EB\ge\left|\br E\times\br B\right|$ that $\left|\br V_{\!\text{A}}\right|\le c$. 

\textbf{Argument from Analogy:} The consensus definition that energy
flow velocity is simply $\br V_{\!\text{A}}=\br S/{\cal E}$ is suggested
by analogy with the well understood example of $\br V_{\!\text{q}}=\br J/\rho$
as the velocity of charge flow, given electric charge density $\rho$
and charge flux density $\br J$. It will therefore be useful to begin
with a review of the properties of charge flow.

The charge flux four-vector is $\fv J=c\rho\fv e_{0}+\br J$. This
four-vector can be timelike, spacelike, or null; but there are useful
cases in which it is timelike. In these cases, the velocity $\br V_{\!\text{qA}}=\br J/\rho$
has magnitude less than the speed of light. 

As noted in \prettyref{sec:Intro}, there are actually two possible
definitions of the velocity of charge flow. One is the $\br V_{\!\text{qA}}=\br J/\rho$
just defined. The other is the velocity $\br V_{\!\text{qB}}$ of
a comoving reference system (which we will refer to as the primed
system) in which the charge flux vector vanishes, $\br J'=0$. Since
an observer at rest in the prime system sees a zero charge flux, this
observer must be moving with the charge flow, and his velocity $\br V_{\!\text{qB}}$
must be the velocity of that flow. 

Since $\fv J=c\rho\fv e_{0}+\br J$ is known to be a four-vector,
Appendix{\,\,}II shows that a Lorentz boost transformation\footnote{The Lorentz boost formalism is summarized in Appendices{\,\,}I and
I.1. } with the boost velocity $\br V=\br V_{\!\text{qA}}=\br J/\rho$ actually
\emph{also} transforms from the original unprimed reference system
to a reference system in which $\br J'=0$. Thus, for the case of
charge flow, the two possible flow velocity definitions coincide,
$\br V_{\!\text{qB}}=\br V_{\!\text{qA}}$.

But the argument in Appendix{\,\,}II depends essentially on the four-vector
transformation rules of $\fv J$. It is crucial that $\fv J=c\rho\fv e_{0}+\br J$
is a legitimate four-vector whose components transform according to
the standard rule $J'^{\alpha}=\Lambda_{\:\,\beta}^{\alpha}\,J^{\beta}$.

If we try to apply the reasoning of Appendix{\,\,}II with $\rho$,
$\br J$ replaced by ${\cal E}$, $\br S$, the chain of logic does
not go through to its conclusion. Unlike $\rho$ and $\br J$, the
energy density ${\cal E}$ and energy-flux vector $\br S$ of the
Poynting theorem do \emph{not }transform as components of a four-vector.
The transformation rule for $c{\cal E}$ and $\left(\br S\right)_{i}$
is as the $\left(00\right)$ and $\left(0i\right)$ components of
the more complicated expression given in \prettyref{eq:cavA2} below,
which will also involve contributions from $cM_{ij}$ terms. Instead
of $J'^{\alpha}=\Lambda_{\:\,\beta}^{\alpha}\,J^{\beta}$ we have
$S'^{\alpha}\neq\Lambda_{\:\,\beta}^{\alpha}\,S^{\beta}$. The argument
in Appendix{\,\,}II thus fails when $\rho$, $\br J$ are replaced
by ${\cal E}$, $\br S$. The boost velocity $\br V_{\!\text{B}}$
that would make $\br S'=0$ does not coincide with $\br V_{\!\text{A}}$. 

The analogy between charge flow and energy flow is therefore broken,
and cannot be used as an argument for the consensus velocity definition
$\br V_{\!\text{A}}=\br S/{\cal E}$.

In the case of the Poynting theorem, $\br V_{\!\text{B}}\neq\br V_{\!\text{A}}$
and both cannot be correct. A choice must be made between them. It
will be shown in \prettyref{sec:CaveatsA} that energy flow velocity
$\br V_{\!\text{A}}$ is inconsistent with the transformation rules
of special relativity and must be rejected. A correct and relativistically
legitimate velocity $\br V_{\!\text{B}}$ is derived in \prettyref{sec:DetailB}.

\textbf{Argument from Geometry:} In addition to the above analogy
with charge density, the following simple geometric construction can
be used to argue for the consensus definition $\br V_{\!\text{A}}=\br S/{\cal E}$. 

\textbf{\emph{Geometry of a Flow:}}\textbf{\emph{\small{} }}\emph{Given
a flowing substance with density $\kappa$ and flux density vector
$\br K$, define a velocity as $\br v=\br K/\kappa$.}

\emph{Now we must examine this velocity definition $\br v$ to see
whether it passes the test of compatibility with the rules of special
relativity. If it does not, then application of the results of this
inset will be a misapplication, and would lead to results inconsistent
with the special theory of relativity. The test must be applied on
a case-by-case basis. Some applications will be seen to be correct,
but others will be misapplications.}

\emph{Assuming that $\br v$ passes that test, the following simple
geometric argument may be made. Consider an arbitrarily oriented area
element $d\br a$ and a time increment $dt.$ The product $d\tau=\left(\br v\,dt\right)\cdot d\br a$
is a volume element. All points in $d\tau$ moving with velocity $\br v$
will flow through $d\br a$ in time $dt$. Now multiply by $\kappa$
to obtain $\kappa d\tau=\kappa\br v\cdot d\br a\,dt$, the amount
of substance in $d\tau$. If we assume that all of the substance in
$d\tau$ is moving with the same velocity $\br v$, then $\br v\cdot d\br a\,dt$
is the amount of substance flowing through $d\br a$ in time $dt$.
But this amount is also, by definition of the flux density $\br K$,
given by $\br K\cdot d\br a\,dt$. Thus 
\begin{equation}
\br K\cdot d\br a\,dt=\kappa\br v\cdot d\br a\,dt\label{eq:geo1}
\end{equation}
Since $d\br a$ and $dt$ are arbitrary, it follows that $\br K=\kappa\br v$.}

\emph{But the above assumption that all elements of the substance
are moving with the same velocity $\br v$ is often unjustified. (Think
of a flow of electrical charges with some thermal velocity.) Then
the above simple geometric argument fails.}

\emph{But if the argument in \prettyref{eq:geo1} fails (while still
assuming that $\br v$ passes the relativity test above) we can consider
$\br v=\br K/\kappa$ to be a }definition \emph{of an average flow
velocity. Then $\br K=\kappa\br v$ is true by definition.}

If this geometrical argument with $\kappa={\cal E}$, $\br K=\br S$
passed the test of compatibility with relativity, it would predict
that the energy flow velocity must be $\br v=\br S/{\cal E}=\br V_{\!\text{A}}$,
the consensus definition of energy flow velocity defined in \prettyref{eq:defA1},
either as a geometric construction or as a definition. However, it
will be shown in the next section that $\br v=\br S/{\cal E}=\br V_{\!\text{A}}$
does not pass the test of compatibility with relativity. 

\section{Caveats of Definition A\label{sec:CaveatsA} }

The principal difficulty with Definition A for the energy flow velocity
is that it is inconsistent with the transformation rules of special
relativity. 

We take a flow velocity definition to be \emph{relativistically valid}
only if that definition passes a simple test using the Einstein velocity
addition formula. Since that test is derived directly from the transformation
rules of special relativity, any flow velocity definition that fails
the test must also violate some rule of special relativity. Definition
A fails this simple test and hence is not relativistically valid.

\textbf{\textit{Einstein Addition Test:}}\textit{ Consider two different
but parallel initial reference systems referred to as the unprimed
and asterisk systems. Let the asterisk system be obtained from the
unprimed one by a boost transformation with boost velocity $\br V=V\fv e_{1}$
of arbitrary magnitude $V$. Then unit vectors $\fv e_{i}$ are parallel
to the corresponding $\fv e_{i}^{*}$, for $i=1,2,3$. Let the velocity
of a comoving observer moving with the energy flow be $\br v$ relative
to the unprimed system, and the velocity of that same observer relative
to the asterisk system be $\br v^{*}$. With no loss of generality,
the unprimed system can be oriented so that $\br v$ is in the $\fv e_{1}$
direction. Then $\br v=v\fv e_{1}$ and hence $\br v^{*}=v^{*}\fv e_{1}^{*}$.}

\textit{Electrodynamics in vacuum except for a possible explicit source
can be expressed in manifestly covariant form and therefore must be
true regardless of the choice of initial reference system; there can
be no privileged initial reference system. So any relativistically
correct derivation that defines flow velocity $\br v$ when applied
in the unprimed system can also be applied to define flow velocity
definition $\br v^{*}$ when applied in the asterisk system. And these
two velocities are velocities of the same comoving observer. Thus,
when we make a boost transformation with boost velocity $\br V=V\fv e_{1}$
between the unprimed and asterisk systems, these velocities $\br v$
and $\br v^{*}$ must transform by the Einstein velocity-addition
formula (See Example 12.6 of Griffiths\cite{Griffiths})
\begin{equation}
\left(v/c\right)=\dfrac{\left(V/c\right)+\left(v^{*}/c\right)}{1+\left(V\,v^{*}/c^{2}\right)}\label{eq:cavA6}
\end{equation}
This is the Einstein Addition Test that any relativistically valid
flow velocity definition must pass.}

\textit{\emph{Note that this test is a necessary condition for consistency
with the transformation rules of special relativity. Any electromagnetic
energy flow definition that fails to pass the }}\textit{Einstein Addition
Test }\textit{\emph{in \prettyref{eq:cavA6} must necessarily violate
the transformation rules of special relativity from which \prettyref{eq:cavA6}
is directly derived, and therefore must be rejected.}}

\textit{\emph{We now show that the consensus energy-flow velocity
definition }}$\br V_{\!\text{A}}=\br S/{\cal E}$ fails the\textit{
Einstein Addition Test}. This failure is due to the fact that $c{\cal E}$
and $\br S$ are not components of a four-vector. Unlike the legitimate
four-vector of charge flow $\fv J=c\rho\fv e_{0}+\br J$, there is
no four-vector $\fv S\asymp c{\cal E}\fv e_{0}+\br S$. (I use the
symbol $\asymp$ to remind the reader that, though written formally
as a four-vector here, it does not actually transform as one.)\footnote{A related point is made by Rohrlich \cite{Rohrlich-answer}, using
the so-called von Laue's theorem to argue that \emph{integrals} of
$c{\cal E}$ and $\br S$ over hyperplanes may in some cases transform
as four-vectors. But we are treating these quantities locally, at
a particular event. Von Laue's theorem does not imply that the local
field functions $c{\cal E}$ and $\br S$ (the integrands of these
hyperplane integrals) themselves transform as components of a four-vector.
They do not. See also Chapter 6 of Rohrlich \cite{Rohrlich}.} Noting that $\br S=c^{2}\br G$ where $\br G$ is the linear momentum
density of the electromagnetic field, the $c{\cal E}$ and $\left(\br S\right)_{i}$
actually transform as the $(00)$ and $(0i)$ components of a four-tensor
$cT^{\alpha\beta}$ defined as $c$ times the standard electromagnetic
energy-momentum tensor 
\begin{equation}
\left(T^{\alpha\beta}\right)=\left(\begin{array}{cccc}
{\cal E} & c\,G_{1} & c\,G_{2} & c\,G_{3}\\
c\,G_{1} & M_{11} & M_{12} & M_{13}\\
c\,G_{2} & M_{21} & M_{22} & M_{23}\\
c\,G_{3} & M_{31} & M_{32} & M_{33}
\end{array}\right)\label{eq:cavA1}
\end{equation}
where $M_{ij}=-\left(E_{i}E_{j}+B_{i}B_{j}\right)\,+\,\dfrac{1}{2}\,\delta_{ij}\,\left(E^{2}+B^{2}\right)$.
Thus the transformation rule for $c{\cal E}$ and $\left(\br S\right)_{i}$
is as the $\left(00\right)$ and $\left(0i\right)$ components of
the more complicated expression 
\begin{equation}
c{T'}^{\alpha\beta}=\Lambda_{\:\,\mu}^{\alpha}\Lambda_{\:\,\nu}^{\beta}cT^{\mu\nu}\label{eq:cavA2}
\end{equation}
 which will also involve contributions from the $cM_{ij}$ terms.

To see \textit{\emph{that energy-flow velocity definition }}$\br V_{\!\text{A}}=\br S/{\cal E}$
fails the \textit{Einstein Addition Test}, begin with the example
of the charge flow definition $\br V_{\!\text{qA}}=\br J/\rho$ that
\emph{passes} the test.

Appendix{\,\,}III demonstrates that the coordinate velocities $\br V_{\!\text{qA}}=\br J/\rho$
and $\br V_{\text{qA}}^{*}=\br J^{*}/\rho^{*}$ derived from the charge
density four-vector $\fv J$ \emph{do} pass the \emph{Einstein Addition
Test,} as must be true for any well-defined coordinate velocity. When
the inverse four-vector transformation rule $J^{\alpha}=\underline{\Lambda}_{\:\,\beta}^{\alpha}\,J{}^{*\beta}$
is used to write $J^{0}$ and $J^{i}$ in terms of $\br V$ and the
asterisk system quantities $J^{*0}$ and $J^{*i}$, the result is
the last expression in \prettyref{eq:fv3}, which agrees with the
Einstein velocity addition rule \prettyref{eq:cavA6} and hence with
special relativity. 

However, if we now attempt to apply this same argument to the case
of $\br V_{\!\text{A}}=\br S/{\cal E}$ and $\br V_{\text{A}}^{*}=\br S^{*}/{\cal E}^{*}$,
the argument of Appendix{\,\,}III fails. In this case, the \emph{equality}
$J^{\alpha}=\underline{\Lambda}_{\:\,\beta}^{\alpha}\,J{}^{*\beta}$
is replaced by the \emph{inequality} $S^{\alpha}\neq\underline{\Lambda}_{\:\,\beta}^{\alpha}\,S{}^{*\beta}$
resulting from the failure of $\fv S$ to be a four-vector. Thus,
eqs.(\ref{eq:fv2}, \ref{eq:fv3}) are not true when $\fv J$ is replaced
by $\fv S$, and the argument does not go through to its conclusion.

In place of the equality in \prettyref{eq:fv3} for the charge density
case, in the case of $\br V_{\!\text{A}}$ we have the \emph{inequality}
\begin{equation}
\left(V_{\!\text{A}}/c\right)=\left(S/{\cal E}c\right)\neq\dfrac{\left(V/c\right)+\left(V_{\!\text{A}}^{*}/c\right)}{1+\left(V\,V_{\!\text{A}}^{*}/c^{2}\right)}\label{eq:cavA3}
\end{equation}
where the expression on the extreme right in \prettyref{eq:cavA3}
would be the correct Einstein velocity addition result. Thus $\br V_{\!\text{A}}$
fails the \emph{Einstein Addition Test.}

The inequality in \prettyref{eq:cavA3} can also be derived directly
from the transformation rules for the $E$ and $B$ fields, without
making any reference to the charge flow analogy. Using the same geometry
as in the \emph{Einstein Addition Test} above, and a transformation
rule similar to \prettyref{eq:defB1}, it can be shown that 
\begin{equation}
\left(V_{\!\text{A}}/c\right)=\dfrac{\left(V/c\right)+\left(1+V^{2}/c^{2}\right)\left(V_{\!\text{A}}^{*}/c\right)}{\left(1+V^{2}/c^{2}\right)+2\left(V\,V_{\!\text{A}}^{*}/c^{2}\right)}\neq\dfrac{\left(V/c\right)+\left(V_{\!\text{A}}^{*}/c\right)}{1+\left(V\,V_{\!\text{A}}^{*}/c^{2}\right)}\label{eq:cavA5}
\end{equation}
which corroborates the inequality in \prettyref{eq:cavA3} and shows
again that $\br V_{\!\text{A}}=\br S/{\cal E}$ fails the \emph{Einstein
Addition Test.}

The results of the present Section can be stated as the following
proposition:

\textbf{\emph{Proposition 1:}}\emph{ The consensus definition of energy
flow velocity}
\begin{equation}
\br S={\cal E}\br V_{\!\text{A}}\quad\quad\text{and}\quad\quad\br V_{\!\text{A}}=\br S/{\cal E}\label{eq:cavA7}
\end{equation}
\emph{defines a velocity $\br V_{\!\text{A}}$ that fails the Einstein}
\emph{Addition Test} \emph{and therefore cannot be used as a relativistically
valid definition for the electromagnetic energy flow velocity. }

In spite of its relativistic incorrectness, the definition in \prettyref{eq:cavA7}
might seem to be proved by the geometrical argument in \emph{Geometry
of a Flow} from \prettyref{sec:DetailA}. However, note that the success
of that geometric argument depends essentially on the assumption that
all elements of the flowing substance are moving with the same velocity.
But the Poynting theorem gives us ${\cal E}$ and $\br S$ as \emph{densities
}and not as precise values. There is no reason to suppose that ${\cal E}$
and $\br S$ are densities of a set of elements all moving at exactly
the same velocity. In fact, the relativistic incorrectness of \prettyref{eq:cavA7}
argues that they are not.

In summary, regardless of how it is derived, either from a flawed
analogy with charge flow, or from a misapplication of \emph{Geometry
of a Flow,} the definition $\br V_{\!\text{A}}=\br S/{\cal E}$ of
energy flow velocity violates the transformation rules of special
relativity and is not relativistically valid. It follows that $\br V_{\!\text{A}}$
cannot be the velocity of electromagnetic energy flow in a relativistically
correct theory.

\section{Detail of Definition B\label{sec:DetailB}}

Definition B of the energy flow velocity, denoted $\br V_{\!\text{B}}$,
is the velocity of a comoving observer who measures a zero energy
flux. Expressed in the precise language of Lorentz boost transformations:

\emph{The coordinate velocity of the flow of electromagnetic field
energy at a given event is the velocity $\br V_{\!\text{B}}$ of a
Lorentz boost that transforms the original reference system into a
reference system in which the Poynting energy flux vector is zero
at that event. }

An observer at that event and at rest in this transformed system,
which we call the comoving system and denote by primes, therefore
measures a zero energy flux. The zero flux measurement indicates that
this observer is comoving with the flow of energy. Such an observer
has coordinate velocity $\br V_{\!\text{B}}$ relative to the original
system,\footnote{See Appendix{\,\,}I.2 for a demonstration that any point at rest
in the primed system moves with  coordinate velocity $\br V_{\!\text{B}}$.} and therefore $\br V_{\!\text{B}}$ is the coordinate velocity of
the energy flow at the given event.

The problem is to find this boost velocity $\br V_{\!\text{B}}$.
An analogous problem arises in the generic theory of relativistic
fluid flow.\footnote{Part I, Chapter 2 of Weinberg\cite{weinberg} presents what I will
refer to as a \emph{generic} theory. It assumes only that a fluid
is composed of a countable set of small particles characterized by
their mass $m_{n}$, position $\br x$$_{n}$, and velocity $\br v_{n}$.
Weinberg (\emph{e.g.} his eq.(2.8.1) \emph{et seq}) uses the language
of Dirac delta function densities, but his formulas are easily translated
into more standard density functions.} There a velocity can be defined as $\br V_{\!\text{a}}=\br pc^{2}/e$
analogous to our $\br V_{\!\text{A}}=\br S/{\cal E}=\br Gc^{2}/{\cal E}$.
But, a proof analogous to the proof in \prettyref{sec:CaveatsA} shows
that velocity to be inconsistent with the Einstein velocity relation
of special relativity and hence not a valid definition. In the theory
of fluid flow, there is no other way to derive a flow velocity from
first principles. One solution is simply to \emph{assert }that there
\emph{must be} a primed reference system moving with the flow even
though we have been unable to derive it; to assert that the energy-momentum
tensor in that system must have the isotropic form $\left(X'^{\alpha\beta}\right)=\text{diag}\{\varepsilon',\pi',\pi',\pi'\}$,
where $\varepsilon'$ is an energy density and $\pi'$is a pressure.
This is called the \emph{perfect fluid }model. However it remains
true that the flow velocity and the form of the energy-momentum tensor
are simply asserted rather than derived.\footnote{Weinberg\cite{weinberg} Part I, Chapter 2, Section 10, eq.(2.10.1)
\emph{et seq}. Note that Weinberg introduces the perfect fluid by
saying, \textquotedbl{}A useful approximation is ...\textquotedbl{}
rather than attempting to derive it from his previous work in his
Chapter 2.}

In the electromagnetic case considered in the present paper, however,
the failure of $\br V_{\!\text{A}}$ does not exhaust our ways of
deriving $\br V_{\!\text{B}}$. We can fall back on the rich structure
of the Maxwell equations themselves, which underlie the definition
of the energy-momentum tensor $T^{\alpha\beta}$ and from which it
was derived. Thus in the electromagnetic case we are not reduced to
merely \emph{asserting} the existence of a comoving frame. We can
actually \emph{derive} the boost velocity $\br V_{\!\text{B}}$ and
the form of the energy-momentum tensor in the comoving frame, starting
from first principles.

The rules for transformation of electric and magnetic fields by a
boost with velocity $\br V_{\!\text{B}}$ can be written in a special
relativistically correct but not manifestly covariant form\footnote{\label{fn:cireqSymbol}See Section 11.10 of Jackson \cite{Jackson},
eq.(11.149). The $\circeq$ symbol means that the components of the
three-vector on the left side of this symbol, expressed in the primed
coordinate system, are numerically equal to the corresponding components
of the three-vector on the right side of this symbol, expressed in
the original unprimed system. If $\br a'\circeq\br c$ and $\br b'\circeq\br d$,
it is easily proved that: (a) $(\br a'\times\br b')\circeq(\br c\times\br d)$
and (b) $(\br a'\cdot\br b'){=}(\br c\cdot\br d)$. (c) Also if $\br w'\circeq\br w$
then the magnitudes are equal, $w'=\left|\br w'\right|=\left|\br w\right|=w$.}
\begin{align}
\br E' & \circeq\gamma_{\!\text{B}}\left(\br E+\dfrac{\br V_{\!\text{B}}}{c}\times\br B\right)+\left(1-\gamma_{\!\text{B}}\right)\dfrac{\br V_{\!\text{B}}\left(\br V_{\!\text{B}}\cdot\br E\right)}{V_{\!\text{B}}^{2}}\nonumber \\
\br B' & \circeq\gamma_{\!\text{B}}\left(\br B-\dfrac{\br V_{\!\text{B}}}{c}\times\br E\right)+\left(1-\gamma_{\!\text{B}}\right)\dfrac{\br V_{\!\text{B}}\left(\br V_{\!\text{B}}\cdot\br B\right)}{V_{\!\text{B}}^{2}}\label{eq:defB1}
\end{align}
where the Lorentz factor is $\gamma_{\!\text{B}}=\left(1-V_{\!B}^{2}/c^{2}\right)^{-1/2}$.

The boost velocity $\br V_{\!\text{B}}$ can then be found by writing
\begin{equation}
\br V_{\!\text{B}}=\lambda\,\br V_{\!\text{A}}\label{eq:defB2}
\end{equation}
where $\lambda$ is a rotationally scalar quantity to be determined.
The velocity $\br V_{\!\text{B}}$ will have the same \emph{direction}
as $\br V_{\!\text{A}}$ but not the same \emph{magnitude.} 

Since $\br V_{\!\text{A}}$ and hence $\br V_{\!\text{B}}$ are perpendicular
to both the electric and magnetic fields, it follows that $\left(\br V_{\!\text{B}}\cdot\br E\right)=\left(\br V_{\!\text{B}}\cdot\br B\right)=0$.
Thus, \prettyref{eq:defB1} reduces to\footnote{\label{fn:dots}Note that $\br V'_{\!\text{B}}{\circeq}\br V_{\!\text{B}}$
as defined in Appendix{\,\,}I.1, together with \prettyref{eq:detB3}
and property (b) of the symbol $\circeq$ in footnote \ref{fn:cireqSymbol},
imply that $(\br V'_{\!\text{B}}\cdot\br E'){=}\br V{}_{\!\text{B}}\cdot\gamma\left[\br E+(\br V{}_{\!\text{B}}/c)\times\br B\right]{=}\gamma(\br V{}_{\!\text{B}}\cdot\br E){=}0$.
Similarly, $(\br V'_{\!\text{B}}\cdot\br B'){=}0$.}
\begin{align}
\br E' & \circeq\gamma_{\!\text{B}}\left(\br E+\dfrac{\br V_{\!\text{B}}}{c}\times\br B\right)\nonumber \\
\br B' & \circeq\gamma_{\!\text{B}}\left(\br B-\dfrac{\br V_{\!\text{B}}}{c}\times\br E\right)\label{eq:detB3}
\end{align}
Insert \prettyref{eq:detB3} into the equation for the Poynting vector
in the comoving system, $\br S'=c\br E'\times\br B'$. Using property
(a) of the symbol $\circeq$ from footnote \ref{fn:cireqSymbol} together
with \prettyref{eq:defB2} and then \prettyref{eq:defA1}, Appendix{\,\,}IV
demonstrates that 
\begin{equation}
\br S'=c\br E'\times\br B'\circeq\gamma_{\!\text{B}}^{2}c\left(\br E\times\br B\right)\left(\left(V_{\!\text{A}}/c\right){}^{2}\,\lambda^{2}-2\lambda+1\right)\label{eq:defB4}
\end{equation}

Choose $\lambda$ to solve the quadratic equation 
\begin{equation}
\left(\left(V_{\!\text{A}}/c\right){}^{2}\lambda^{2}-2\lambda+1\right)=0\label{eq:defB5}
\end{equation}
Then \prettyref{eq:defB4} makes $\br S'=0$, as required by Definition
B. The solution is
\begin{equation}
\lambda=\dfrac{1}{\left(V_{\!\text{A}}/c\right){}^{2}}\left\{ 1-\sqrt{1-\left(V_{\!\text{A}}/c\right){}^{2}\,}\,\right\} \label{eq:defB6}
\end{equation}
From \prettyref{eq:defB2}, the Definition B for the velocity of the
energy flow is therefore
\begin{equation}
\br V_{\!\text{B}}=\lambda\,\br V_{\!\text{A}}=\dfrac{1}{\left(V_{\!\text{A}}/c\right){}^{2}}\left\{ 1-\sqrt{1-\left(V_{\!\text{A}}/c\right){}^{2}\,}\,\right\} \br V_{\!\text{A}}\label{eq:defB7}
\end{equation}
where $\br V_{\!\text{A}}$ is defined in \prettyref{eq:defA1}.

This $\br V_{\!\text{B}}$ is the relativistically correct boost velocity
from the original unprimed frame to the comoving reference frame in
which $\br S'=0$.\footnote{Appendix{\,\,}V gives details of the comoving system for possible
values of $(\br E\cdot\br B)$ at a given event.} 

Since $\br V_{\!\text{B}}$ is parallel to the energy flux vector
$\br S$, the energy flow velocity can also be written as $\br V_{\!\text{B}}=V_{\!\text{B}}\left(\br S/S\right)$
where the magnitude $V_{\!\text{B}}$ is given by\footnote{The text just after \prettyref{eq:defA1} proves that $0\le V_{\!\text{A}}\le c$.
As $\left(V_{\!\text{A}}/c\right)$ increases from $0$ to $1$, \prettyref{eq:defB8}
shows that $\left(V_{\!\text{B}}/c\right)$ increases monotonically
from $0$ to $1$, with $V_{\!\text{B}}\le V_{\!\text{A}}$ at every
point. It follows that $0\le V_{\!\text{B}}\le c$ also. Regions of
the unprimed system where ${\cal E}$ is nonzero but $\br S$ is zero
have $V_{\!\text{A}}=0$ and $V_{\!\text{B}}=0$, and have no energy
flow.} 
\begin{equation}
\left(V_{\!\text{B}}/c\right)=\dfrac{1}{\left(V_{\!\text{A}}/c\right)}\left\{ 1-\sqrt{1-\left(V_{\!\text{A}}/c\right){}^{2}\,}\,\right\} \label{eq:defB8}
\end{equation}
Eq.(\ref{eq:defB8}) can be inverted to give 
\begin{equation}
\left(V_{\!\text{A}}/c\right)=\dfrac{2\left(V_{\!\text{B}}/c\right)}{1+\left(V_{\!\text{B}}/c\right)^{2}}\label{eq:defB9}
\end{equation}
which can be used to write the factor $\lambda$ in \prettyref{eq:defB6}
as a function of the velocity Definition B
\begin{equation}
\lambda=\dfrac{1+\left(V_{\!\text{B}}/c\right)^{2}}{2}\label{eq:defB10}
\end{equation}
which shows $\lambda\le1$ and hence $V_{\!\text{B}}\le V_{\!\text{A}}$.

Summary: This section uses the co-variant field transformation equations
in \prettyref{eq:defB1} to derive a boost velocity, $\br V_{\!\text{B}}$,
defined in \prettyref{eq:defB7}, that transforms from the unprimed
system to a comoving primed system in which the energy flux vector
$\br S'=0$. Then Appendix I.2 shows that $\br V_{\!\text{B}}$ is
also the coordinate velocity relative to the unprimed system of an
observer at rest in the comoving primed system. Since it is derived
directly from the rules of special relativity, this velocity is well
defined and relativistically correct. Also, it can be shown that this
$\br V_{\!\text{B}}$ passes the \textit{Einstein Addition Test},
as it must. 

The observer at rest in the primed comoving system will observe the
energy flux vector $\br S'$ to be zero. Thus if he holds an oriented
area element $d\br a'$ in any orientation he will find that the energy
flux through that element to be $\br S'\cdot d\br a'=0$. Hence the
observer must be moving at the same velocity as the flow of energy,
and its velocity will be the same as his velocity, $\br V_{\!\text{B}}$. 

The conclusion is that the well defined and relativistically correct
coordinate velocity $\br V_{\!\text{B}}$ must be the correct velocity
of the electromagnetic energy flow.

This conclusion, together with $\br V_{\!\text{B}}\neq\br V_{\!\text{A}}$
from \prettyref{eq:defB7}, also gives independent conformation of
the results of \prettyref{sec:CaveatsA}, that the correct definition
of electromagnetic energy flow velocity is not the consensus value
$\br V_{\!\text{A}}$. It is important to note that this conclusion,
along with all the results in \prettyref{sec:DetailB}, depend only
on the assumption of the standard transformation laws of electromagnetic
field in \prettyref{eq:defB1}, and not on any other assumptions. 

Thus \prettyref{sec:DetailB} provides convincing proof that $\br V_{\!\text{B}}$
is the relativistically correct electromagnetic energy flow velocity
definition, and that the consensus value $\br V_{\!\text{A}}$ is
not.

\section{Caveats of Definition B\label{sec:CaveatsB}}

The caveats for Definition A are technical; they concern its violation
of the transformation rules of special relativity. By contrast, the
derivation of $\br V_{\!\text{B}}$ in \prettyref{sec:DetailB} is
completely consistent with special relativity throughout. Velocity
$\br V_{\!\text{B}}$ is the relativistically correct velocity of
an observer at rest in the primed comoving reference system, defined
as a system in which the energy flux vector $\br S'=0$. It follows
that $\br V_{\!\text{B}}$ is the relativistically valid energy flow
velocity.

But one may question whether the condition $\br S'=0$ used in \prettyref{sec:DetailB}
truly implies that the comoving observer is moving at the same velocity
as the underlying energy flow, as required for $\br V_{\!\text{B}}$
to be the correct energy flow velocity. For, as \prettyref{eq:defA1}
and \prettyref{eq:defB7} directly prove, $\br V_{\!\text{B}}\neq\br V_{\!\text{A}}=\br S/{\cal E}$
and hence $\br S\neq{\cal E}\br V_{\!\text{B}}$. It may seem that
this inequality will block derivation of the Poynting theorem on which
the meanings of $\br S$ and ${\cal E}$ depend. 

However, the equality of $\br S$ and ${\cal E}\br V_{\!\text{B}}$
is \emph{not} a necessary condition for the Poynting theorem. Derivation
of the Poynting theorem is independent of the relation between $\br S$
and the product ${\cal E}\br V_{\!\text{B}}$. The Poynting conservation
of energy theorem derives from the divergence of the symmetric energy-momentum
tensor $T^{\mu\nu}$ defined in \prettyref{eq:cavA1}
\begin{equation}
\partial_{\mu}T^{\mu\nu}=-f^{\nu}\quad\text{where}\quad f^{\alpha}=\frac{1}{c}F_{\,\,\,\beta}^{\alpha}\,J^{\beta}\label{eq:cb1}
\end{equation}
is the Lorentz force density four-vector and $F^{\mu\nu}$ is the
electromagnetic field tensor.\footnote{See Section 7.3 of Rindler\cite{rindler}.}
The $\nu=0$ component of the above manifestly covariant equation
expands to
\begin{equation}
\dfrac{\partial{\cal E}}{\partial t}+\bg\nabla\cdot\br S=-\br E\cdot\br J\label{eq:cb2}
\end{equation}
which is the Poynting work-energy theorem of electromagnetism. Since
it is derived from the manifestly covariant pair of equations, \prettyref{eq:cb1}
the Poynting energy conservation formula \prettyref{eq:cb2} is well
defined and relativistically correct. And the meanings of ${\cal E}$
and $\br S$ as energy density and energy-flux vector, respectively,
are established by \prettyref{eq:cb2}. No further proof is required.
The Poynting theorem and the meaning of $\br S$ as the energy flux
vector are thus proved, regardless of the relation between $\br S$
and the product ${\cal E}\br V_{\!\text{B}}$. 

This proof that the Poynting theorem and the meaning of the energy
flux vector $\br S$ are independently established corroborates and
completes the argument at the end of \prettyref{sec:DetailB}, which
depended on the meaning of $\br S$. Thus we are driven to the conclusion
that $\br V_{\!\text{B}}$ is indeed the well defined and relativistically
correct velocity of the electromagnetic energy flow, and that $\br V_{\!\text{A}}$
is not.

\section{An Exceptional Case\label{sec:Exceptional}}

Although $\br V_{\!\text{A}}\neq\br V_{\!\text{B}}$ in general, there
is an important exceptional case, which the theory here must approach
as a limit. A plane, monochromatic, right /left circularly polarized
light wave in vacuum with angular velocity $\omega$ and wave vector
$\br k=\left(\omega/c\right)\fv e_{3}$ has 
\begin{align}
\br E & =E_{0}\left\{ \fv e_{1}\,\cos\phi\pm\fv e_{2}\,\sin\phi\right\} \nonumber \\
\br B & =E_{0}\left\{ \mp\fv e_{1}\,\sin\phi+\fv e_{2}\,\cos\phi\right\} \label{eq:why1}
\end{align}
where $\phi=\left(kz-\omega t\right)$ and $z=x^{3}$. This electromagnetic
field has $\br E\perp\br B$ and $E=B=E_{0}\neq0$, which is the limiting
case treated in item (c) of Appendix{\,\,}V. In this exceptional
case, velocity Definitions A and B coincide. As can be seen from \prettyref{eq:defA1}
and \prettyref{eq:defB8} $V_{\!\text{B}}=c=V_{\!\text{A}}$.

As noted in Appendix{\,\,}V, and as also can be read from \prettyref{eq:mom2},
in this case ${\cal E}'$ would be zero in the comoving system. But
there is no comoving system with velocity magnitude equal to the speed
of light. Observers are not permitted to ride on light waves. However,
both definitions do agree that the flow speed of a light wave is the
speed of light.

Setting $V_{\!\text{A}}=c$ and using $\br S=c^{2}\br G$, \prettyref{eq:defA3}
in this special case implies that 
\begin{equation}
Gc={\cal E}\label{eq:why2}
\end{equation}
Since wave solution \prettyref{eq:why1} defines a mode of the electromagnetic
field whose second-quantization creates photons of definite vector
momentum, \prettyref{eq:why2} can be considered a classical precursor
of the relation $pc=e$ for the photon momentum and energy, a relation
that requires the photon to be a massless particle.

\section{The Energy-Momentum Tensor in a Comoving Frame}

\label{sec:EnergyMomentum}The derivation of velocity $\br V_{\!\text{B}}$
in \prettyref{sec:DetailB} also allows the electromagnetic energy-momentum
tensor in the comoving system to be derived from first principles.
As noted in \prettyref{sec:DetailB}, the comoving energy-momentum
tensor of a perfect fluid must simply be asserted rather than derived.
But the electromagnetic energy-momentum tensor in a comoving system
can be \emph{derived,} and shown equal to a simple, diagonal form
depending only on the energy density and one other parameter.

In the comoving (primed) coordinate system that was produced by the
Lorentz boost $\br V_{\!\text{B}}$, the energy-momentum tensor \prettyref{eq:cavA1}
is represented by the tensor components $T'^{\alpha\beta}$ in which
the $cG_{i}'=S'_{i}/c=0$.
\begin{equation}
\left(T'^{\alpha\beta}\right)=\left(\begin{array}{cccc}
{\cal E}' & 0 & 0 & 0\\
0 & M'_{11} & M'_{12} & M'_{13}\\
0 & M'_{21} & M'_{22} & M'_{23}\\
0 & M'_{31} & M'_{32} & M'_{33}
\end{array}\right)\label{eq:mom1}
\end{equation}
where\footnote{Eq.(\ref{eq:mom2}) is derived in Appendix{\,\,}IV. }
\begin{flalign}
{\cal E}' & =\dfrac{1}{2}\left(E'^{2}+B'^{2}\right)={\cal E}\,\dfrac{1-\left(V_{\!\text{B}}/c\right)^{2}}{1+\left(V_{\!\text{B}}/c\right)^{2}}\label{eq:mom2}\\
\text{and}\quad M'_{ij} & =-\left(E'_{i}E'_{j}+B'_{i}B'_{j}\right)+\delta_{ij}{\cal E}'\nonumber 
\end{flalign}

We can now make another Lorentz transformation, an orthogonal spatial
rotation at fixed time, to diagonalize the real, symmetric sub-matrix
$M_{ij}'$ in \prettyref{eq:mom1}. 

The required spatial rotation can be defined as the product of two
proper rotations. First, rotate the coordinate system to bring the
$\fv e'_{3}$ axis into the $\br V_{\!\text{B}}'\circeq\br V_{\!\text{B}}$
direction.\footnote{Note that item (c) of footnote \ref{fn:cireqSymbol} implies equal
magnitudes $V_{\!\text{B}}'=V_{\!\text{B}}$.} Denote this rotated system by tildes. Rotations do not change three-vectors,
which are invariant objects under rotations. However, rotations do
change the \emph{components} of three-vectors. Thus $\tilde{\br V}_{\!\text{B}}{=}\br V_{\!\text{B}}'$,
$\,\tilde{\,\br E}{=}\br E'$, and $\tilde{\br B}{=}\br B'$, but
in the tilde system $\tilde{\br V}_{\!\text{B}}$ now has components
$\tilde{V}_{\!\text{B}1}=\tilde{V}_{\!\text{B}2}=0$ and $\tilde{V}_{\!\text{B}3}=V_{\!\text{B}}$.
Then using footnote \ref{fn:dots}, we have $0=(\br E'\cdot\br V_{\!\text{B}}')=(\tilde{\br E}\cdot\tilde{\br V}_{\!\text{B}})=V_{\!\text{B}}\tilde{E}_{3}$.
Except in no-flow regions with ${\cal E}$ nonzero but $\br S$ zero,
the magnitude $V_{\!\text{B}}\neq0$ and thus  $\tilde{E}_{3}=0$.
A similar argument proves that $\tilde{B}_{3}=0$. Thus the $\left(33\right)$
component of the energy-momentum tensor when expressed in the tilde
system is $\tilde{T}^{33}=-\left({\tilde{E}_{3}}^{2}+{\tilde{B}_{3}}^{2}\right)+\tilde{{\cal E}}=\tilde{{\cal E}}$.
The tensor from \prettyref{eq:mom1}, when expressed in the tilde
system, becomes
\begin{equation}
\left(\tilde{T}^{\alpha\beta}\right)=\left(\begin{array}{cccc}
\tilde{{\cal E}} & 0 & 0 & 0\\
0 & \tilde{M}_{11} & \tilde{M}_{12} & 0\\
0 & \tilde{M}_{21} & \tilde{M}_{22} & 0\\
0 & 0 & 0 & \tilde{{\cal E}}
\end{array}\right)\label{eq:mom3}
\end{equation}
where $\tilde{{\cal E}}={\cal E}'$. 

Since the invariant trace of the electrodynamic energy-momentum tensor
vanishes,\footnote{See Section 7.8 of Rindler \cite{rindler}} it
follows from \prettyref{eq:mom3} that 
\begin{equation}
0=\eta_{\alpha\beta}\tilde{T}^{\alpha\beta}=-\tilde{{\cal E}}+\tilde{M}_{11}+\tilde{M}_{22}+\tilde{{\cal E}}\label{eq:mom4}
\end{equation}
and hence $\tilde{M}_{11}=-\tilde{M}_{22}$. Also, the symmetry of
the energy-momentum tensor makes $\tilde{M}_{21}=\tilde{M}_{12}$.
Thus 
\begin{equation}
\left(\tilde{T}^{\alpha\beta}\right)=\left(\begin{array}{cccc}
\tilde{{\cal E}} & 0 & 0 & 0\\
0 & -\tilde{\psi} & \tilde{\xi} & 0\\
0 & \tilde{\xi} & \tilde{\psi} & 0\\
0 & 0 & 0 & \tilde{{\cal E}}
\end{array}\right)\label{eq:mom5}
\end{equation}
where $\tilde{\psi}=\tilde{M}_{22}$ and $\tilde{\xi}=\tilde{M}_{12}$.

A second proper rotation, this time about the $\tilde{\fv e}_{3}$
axis, produces the final coordinate system, denoted with double primes.
After this rotation, $E''_{3}=\tilde{E}_{3}=0$, $B''_{3}=\tilde{B}_{3}=0$,
and $\br V_{\!\text{B}}''=\tilde{\br V}_{\!\text{B}}$ has components
$V''_{\!\text{B}1}=V''_{\text{B}2}=0$ and $V''_{\text{B}3}=V_{\!\text{B}}$.
The only effect of this second rotation is to diagonalize the 2x2
matrix $\left(\begin{array}{cc}
-\tilde{\psi} & \tilde{\xi}\\
\tilde{\xi} & \tilde{\psi}
\end{array}\right)$. The energy-momentum tensor then has its final diagonal form in the
double-prime system
\begin{equation}
{T''}^{\alpha\beta}=\left(\begin{array}{cccc}
{\cal E}'' & 0 & 0 & 0\\
0 & -a'' & 0 & 0\\
0 & 0 & a'' & 0\\
0 & 0 & 0 & {\cal E}''
\end{array}\right)\label{eq:mom6}
\end{equation}
where ${\cal E}''=\tilde{{\cal E}}={\cal E}'$. The parameter $a''$
has absolute value $\left|a''\right|=\left\{ \tilde{\psi}^{2}+\tilde{\xi}^{2}\right\} ^{1/2}$where
$\pm\left\{ \tilde{\psi}^{2}+\tilde{\xi}^{2}\right\} ^{1/2}$ are
the two eigenvalues of the matrix $\left(\begin{array}{cc}
-\tilde{\psi} & \tilde{\xi}\\
\tilde{\xi} & \tilde{\psi}
\end{array}\right)$ that were calculated during the diagonalization process. The sign
of $a''$ depends on the directions and relative magnitudes of the
electric and magnetic fields. 

The rotation that takes the system from the primed to the double-primed
system is then the product of the first and second rotations. The
various representations of the boost velocity used above are related
by $\br V_{\!\text{B}}''=V_{\!\text{B}}\fv e''_{3}=\tilde{\br V}_{\!\text{B}}=V_{\!\text{B}}\tilde{\fv e}_{3}=\br V_{\!\text{B}}'\circeq\br V_{\!\text{B}}$.
It follows from item (c) of footnote \ref{fn:cireqSymbol} that all
of these vectors have the same original magnitude $V_{\!\text{B}}$.

The energy-momentum tensor \prettyref{eq:mom6} in the double-prime
system is diagonal and in a canonical form, with two elements equal
to ${\cal E}''={\cal E}'$ and two other elements equal to plus or
minus the single parameter $a''$. 

The reduction of the electromagnetic energy-momentum tensor to the
diagonal form in \prettyref{eq:mom6} has important consequences for
possible fluid-dynamic models of electromagnetic energy flow. For
example, the perfect fluid model\footnote{See Part I, Chapter 2, Section 10, eq.(2.10.1) \emph{et seq} of Weinberg
\cite{weinberg}.} has a comoving energy-momentum tensor given by the diagonal matrix
$\left(X'^{\alpha\beta}\right)=\text{diag}\left(\varepsilon',\pi',\pi',\pi'\right)$
where $\varepsilon'$ is an energy density and the $\pi'$ are isotropic
pressure terms, all of which are equal by definition. But, regardless
of the value of parameters ${\cal E}''$ and $a''$, there is no choice
of $\varepsilon'$ and $\pi'$ for which the quadruplet of numbers
$\left(\varepsilon',\pi',\pi',\pi'\right)$ can match the quadruplet
of numbers $\left({\cal E}'',-a'',a'',{\cal E}''\right)$, other than
the unphysical case when all of the numbers in both quadruplets are
zero. Similarly, the so-called dust model\emph{}\footnote{Discussed in\emph{ }Section 12.2 of d'Inverno \cite{dinverno} and
on page 301 \emph{et seq. }of Rindler \cite{rindler}.} has $\left(X'^{\alpha\beta}\right)=\text{diag}\left(\varepsilon',0,0,0\right)$
which also cannot match the electromagnetic tensor.

Hence, the energy-momentum tensor of electrodynamics cannot be successfully
modeled with either a perfect-fluid or a dust model.

\section{Conclusion\label{sec:Conclusion}}

The title of this paper asks whether electromagnetic field momentum
is due to the flow of field energy. The answer has required careful
examination of the velocity of energy flow.

First, consider velocity Definition A from \prettyref{sec:DetailA}.
Dividing the definition $\br S={\cal E}\br V_{\!\text{A}}$, from
\prettyref{eq:defA3} by $c^{2}$ and using $\br G=\br S/c^{2}$ and
${\cal M}_{\text{rel}}={\cal E}/c^{2}$ gives
\begin{equation}
\br G={\cal M}_{\text{rel}}\br V_{\!\text{A}}\label{eq:con1}
\end{equation}
which exhibits electromagnetic field momentum density $\br G$ as
due to the flow of relativistic mass-energy ${\cal M}_{\text{rel}}$.
But as shown in \prettyref{sec:CaveatsA}, $\br V_{\!\text{A}}=\br S/{\cal E}$
is not a relativistically correct velocity definition, and therefore
must be rejected.

Definition B, on the other hand, is derived in \prettyref{sec:DetailB}
with complete adherence to the transformation rules of special relativity.
Definition B is derived from the condition that the coordinate velocity
of the flow of electromagnetic field energy at a given event is the
velocity $\br V_{\!\text{B}}$ of a Lorentz boost that transforms
the original reference system into a reference system in which the
Poynting energy flux vector is zero at that event.\emph{ }But Definition
B does not permit the explanation of field momentum density as due
to moving relativistic mass-energy. Introducing $\br V_{\!\text{B}}=\lambda\,\br V_{\!\text{A}}$
from \prettyref{eq:defB2} into \prettyref{eq:con1} and using \prettyref{eq:defB10}
gives 
\begin{equation}
\br G=\dfrac{{\cal M}_{\text{rel}}}{\lambda}\left(\lambda\br V_{\!\text{A}}\right)=\dfrac{{\cal M}_{\text{rel}}}{\lambda}\br V_{\!\text{B}}=\dfrac{2{\cal M}_{\text{rel}}\br V_{\!\text{B}}}{1+\left(V_{\!\text{B}}/c\right)^{2}}\label{eq:con2}
\end{equation}

Flow of relativistic mass ${\cal M}_{\text{rel}}$ at velocity $\br V_{\!\text{B}}$
would produce a momentum density ${\cal M}_{\text{rel}}\br V_{\!\text{B}}=\lambda\br G$
that has the same direction as $\br G$ but has a magnitude that is
too small by the factor $\lambda\le1$ defined in eqs.(\ref{eq:defB6},
\ref{eq:defB10}).

Note that this failure of the flow of relativistic mass ${\cal M}_{\text{rel}}$
to explain the field momentum density $\br G$ in the electromagnetic
fields \emph{must not be confused with the so-called hidden momentum
}in the sources that is sometimes invoked to balance the field momentum
and preserve momentum conservation globally.\footnote{See Example 12.12 of Griffiths \cite{Griffiths}, and also McDonald
\cite{McDonald-CinBfield} and Babson et al \cite{Babson-hiddenMom}.} 

The present paper is concerned only with a correct understanding of
the electromagnetic \emph{field contribution itself,} locally at every
point of the electromagnetic field including those points with no
source density. Encouraged by the arguments from the Feynman example
noted in footnote \ref{fn:Feynman-et-al} above, we accept that the
vector $\br G=\br S/c^{2}$ correctly reproduces the local field momentum
density at every point of the electromagnetic field. The question
is the source of that local point-by-point field momentum density. 

The conclusion of this paper can be now stated as the following proposition:

\textbf{\emph{Proposition 2:}}\emph{ There is no  relativistically
correct definition of energy flow velocity that explains the electromagnetic
field momentum density as due to the flow of field energy.}

\textbf{\emph{Proof:}}\emph{ If an energy flow velocity $\br v$ obeys
}$\br G={\cal M}_{\text{rel}}\br v$,\emph{ then multiplying by $c^{2}$
implies $\br S={\cal E}\br v$ which in turn implies that $\br v=\br S/{\cal E}=\br V_{\!\text{A}}$.
But, according to Proposition 1 in \prettyref{sec:CaveatsA}, $\br V_{\!\text{A}}$
is not a relativistically correct velocity for electromagnetic energy
flow. This $\br v$ is therefore not relativistically correct. }

The title question of this paper has a negative answer. When adherence
to the strict transformation rules of special relativity is required,
electromagnetic field momentum cannot be explained as due to the flow
of field energy. 

\section{Afterword\label{sec:Maxwell}}

Detailed studies of the energy and momentum carried by the electromagnetic
field, such as the present paper, can be seen as searches for clues
to a possible new physics underlying the Maxwell Equations. But, if
we accept the conclusion at the end of \prettyref{sec:Conclusion},
the attempt to model the Maxwell Equations at the level of energy
flow and the energy-momentum tensor seems a failed program. 

This failure calls into question the whole project of finding a deeper
level behind the Maxwell Equations. A consensus exists that the electric
and magnetic fields are not states of anything else,\footnote{See, for example, Section 4-5 of Feynman \cite{FeynmanLectures}.}
but are either abstract mathematical aids, or themselves elements
of reality to be taken as fundamental. In this view, the Maxwell Equations
are already at the fundamental level, and attempts to derive them
from some deeper reality are a futile revival of nineteenth century
aether theories and, as Feynman says, \textquotedbl{}... produce nothing
but errors.\textquotedbl{}

But Maxwell himself looked for fluid models of his equations. Maxwell
\cite{Maxwell-incompressible} explains the inverse square electric
force law as a consequence of the spread of an incompressible fluid.
And he later proposes (Maxwell \cite{Maxwell-vortex}) a model of
Faraday's magnetic field lines based on fluid vortices.\footnote{Falconer \cite{Falconer} surveys other early vortex models.} 

Perhaps, instead of taking the conclusion in \prettyref{sec:Conclusion}
as a reason to abandon Maxwell's search, we should rather read a lesson
from it: \emph{Our attempt at a flow model may have failed because
the attempt is taking place at the wrong level.} The electromagnetic
energy-momentum tensor $T^{\alpha\beta}$ is quadratic in the fundamental
electromagnetic fields $E$ and $B$. It may be that any successful
flow model of electrodynamics must operate at the linear, field level
and not at the energy-momentum level.

For example, consider two monochromatic plane waves propagating in
the $+\fv e_{3}$ direction, wave $a$ with right circular polarization
and wave $b$ with left circular polarization.
\begin{multline}
\br E_{a}=E_{0}\left\{ \fv e_{1}\,\cos\phi+\fv e_{2}\,\sin\phi\right\} \\
\br E_{b}=E_{0}\left\{ \fv e_{1}\,\cos\phi-\fv e_{2}\,\sin\phi\right\} \label{eq:aft1}
\end{multline}
where $\phi=\left(kz-\omega t\right)$and $z=x^{3}$, and the magnetic
fields are the cross product of $\fv e_{3}$ with the given electric
fields. The electromagnetic energy-momentum tensors $T_{a}^{\alpha\beta}$
and $T_{b}^{\alpha\beta}$ of these two waves are the same, with 
\begin{equation}
\left(T_{a}^{\alpha\beta}\right)=\left(T_{b}^{\alpha\beta}\right)=\left(\begin{array}{cccc}
E_{0}^{2} & 0 & 0 & E_{0}^{2}\\
0 & 0 & 0 & 0\\
0 & 0 & 0 & 0\\
E_{0}^{2} & 0 & 0 & E_{0}^{2}
\end{array}\right)\label{eq:aft2}
\end{equation}

We now want to superpose these two situations $a$ and $b$. The superposition
of the two circularly polarized waves is a linearly polarized wave
\begin{equation}
\br E_{a+b}=\br E_{a}+\br E_{b}=2E_{0}\fv e_{1}\cos\phi\label{eq:aft3}
\end{equation}
and the resulting energy-momentum tensor is 
\begin{equation}
\left(T_{a+b}^{\alpha\beta}\right)=\left(\begin{array}{ccccc}
4E_{0}^{2}\cos^{2}\phi & 0 &  & 0 & 4E_{0}^{2}\cos^{2}\phi\\
0 & 0 &  & 0 & 0\\
0 & 0 &  & 0 & 0\\
4E_{0}^{2}\cos^{2}\phi & 0 &  & 0 & 4E_{0}^{2}\cos^{2}\phi
\end{array}\right)\label{eq:aft4}
\end{equation}
which is time varying at each fixed spatial point, passing\linebreak 
through zero every $\pi/\omega$ seconds.

This example illustrates that representing the electromagnetic energy-momentum
flow as the flow of a fluid at the quadratic energy-momentum level
ignores the fact that electromagnetism is a linear theory with superposition.
It is difficult to see how combining the two tensors in \prettyref{eq:aft2}
could result in the time-varying tensor of \prettyref{eq:aft4}. Electromagnetic
fields do not superpose at the energy-momentum level. Therefore an
attempt to model electromagnetism at that level is bound to fail.
Such a model should be applied at the linear, field level of the $E$
and $B$ fields themselves.

But, in spite of the appeal and long history of Maxwell's quest, there
are formidable hurdles facing any model at the field level, even using
modern mathematical techniques. One such hurdle is that a complete
model of the $E$ and $B$ fields would probably also need to include
interaction with, and characterization of, the source fields $\rho$
and $\br J$. And it should include not only the effects of sources
on fields but also the effects of fields on sources, the Lorentz force
law.

\section*{Appendix I: Lorentz Boosts\label{sec:I-Boosts}}

Consider a Lorentz transformation from an \textquotedbl{}unprimed\textquotedbl{}
coordinate system  with coordinates $x=(x^{0},x^{1},x^{2},x^{3})$
to a \textquotedbl{}primed\textquotedbl{} coordinate system with coordinates
$x'=(x'^{0},x'^{1},x'^{2},x'^{3})$ where $x^{0}=ct$ and $x'^{0}=ct'$.
The most general proper, homogeneous Lorentz transformation from the
unprimed to the primed systems can be written as a Lorentz boost times
a rotation.\footnote{See Part I, Chapter 2, Section 1 of Weinberg \cite{weinberg}.} 

\subsection*{I.1: Definition of Lorentz Boost}

A Lorentz boost transformation is parameterized by a boost velocity
vector $\br V$ with components $(V_{1},V_{2},V_{3})$ and magnitude
$V=\left(V_{1}^{2}+V_{2}^{2}+V_{3}^{2}\right)^{1/2}$. Using the Einstein
summation convention, it is written as $x'^{\alpha}=\Lambda_{\:\,\beta}^{\alpha}\,x^{\beta}$
where $\Lambda_{\:\,0}^{0}=\gamma$,~ $\Lambda_{\:\,i}^{0}=\Lambda_{\:\,0}^{i}=-\gamma V_{i}/c$,
and $\Lambda_{\:\,j}^{i}=\delta_{ij}+(\gamma-1)V_{i}V_{j}/V^{2}$.
The $\delta_{ij}$ is the Kroeneker delta function, and $\gamma=\left(1-V^{2}/c^{2}\right)^{-1/2}$. 

The inverse boost $\underline{\Lambda}_{\:\,\beta}^{\alpha}$ is the
same except for the substitution $V_{i}\rightarrow-V_{i}$. Thus the
inverse boost vector is $\left(-\br V'\right)$ where $\br V'\circeq\br V$.
(See footnote \ref{fn:cireqSymbol} for definition of the $\circeq$
symbol.)

\subsection*{I.2: Meaning of the Boost Velocity $\br V$}

The velocity $\br V$ that parameterizes the Lorentz boost is also
the coordinate velocity, as measured from the unprimed system, of
any point that is at rest in the primed system. In this sense, the
entire primed system is moving with velocity $\br V$ as observed
from the unprimed system. Any observer at rest in the primed system
is moving with that velocity $\br V$ relative to the unprimed system.

To see this, apply the inverse Lorentz boost to the differentials
of a point at rest in the primed system, $dx'^{i}=0$ for $i=1,2,3$,
but $dx'^{0}>0$. The result is $dx^{0}=\gamma dx'^{0}$ and $dx^{i}=\gamma\left(V_{i}/c\right)dx'^{0}$.
Thus $dx^{i}/dt=V_{i}$, as was asserted.

\section*{Appendix II: Proof that Boost with $\br V_{\!\text{qA}}$ Makes $\br J'=0$}

As applied to a four-vector $\fv J=J^{0}\fv e_{0}+\br J$, with $J^{0}=c\rho$
and $J^{i}=\left(\br J\right)_{i}$ the Lorentz boost transformation
rule is $J'^{\alpha}{=}\Lambda_{\:\,\beta}^{\alpha}\,J^{\beta}$.
Hence
\begin{multline}
J'^{i}=\Lambda_{\:\,0}^{i}J^{0}+\Lambda_{\:\,j}^{i}J^{j}\\
=-\gamma\dfrac{V_{i}}{c}J^{0}+J^{i}+(\gamma-1)\dfrac{V_{i}\left(V_{j}J^{j}\right)}{V^{2}}\label{eq:fv1}
\end{multline}
Replacing boost velocity ratio $V_{i}/c$ by $\left(\br V_{\!\text{qA}}\right)_{i}/c=J^{i}/J^{0}$
in \prettyref{eq:fv1} makes $J'^{i}=0$, as asserted.

\section*{Appendix III: Proof that $\br V_{\!\text{qA}}$ is Consistent with
the Einstein Velocity Addition Formula}

Consider two reference frames, one denoted as unprimed and the other
with asterisks. Let the orientation of the frames be as in the \emph{Einstein
Addition Test }in \prettyref{sec:CaveatsA}.

The four-vector charge flux in the asterisk system is then $\fv J=J^{*0}\fv e_{0}^{*}+J^{*1}\fv e_{1}^{*}$
where $J^{*0}=c\rho^{*}$ and $J^{*1}=\left(\br J^{*}\right)_{1}$.
From the standard inverse boost formula $J^{\alpha}=\underline{\Lambda}_{\:\,\beta}^{\alpha}\,J{}^{*\beta}$,
the transform between the two frames is (suppressing the 2 and 3 components
for simplicity)
\begin{equation}
\left(\begin{array}{c}
J^{0}\\
J^{1}
\end{array}\right)=\gamma\left(\begin{array}{cc}
1 & \left(V/c\right)\\
\left(V/c\right) & 1
\end{array}\right)\left(\begin{array}{c}
J^{*0}\\
J^{*1}
\end{array}\right)\label{eq:fv2}
\end{equation}
The velocities $V_{\!\text{qA}}$ and $V_{\!\text{qA}}^{*}$ in the
two frames are therefore related by 
\begin{multline}
\frac{V_{\!\text{qA}}}{c}=\dfrac{J^{1}}{J^{0}}=\dfrac{\left(V/c\right)J^{*0}+J^{*1}}{J^{*0}+\left(V/c\right)J^{*1}}\\
=\dfrac{\left(V/c\right)+\left(J^{*1}/J^{*0}\right)}{1+\left(V/c\right)\left(J^{*1}/J^{*0}\right)}=\dfrac{\left(V/c\right)+\left(V_{\!\text{qA}}^{*}/c\right)}{1+\left(V\,V_{\!\text{qA}}^{*}/c^{2}\right)}\label{eq:fv3}
\end{multline}
 which replicates the standard Einstein velocity addition formula,
as asserted. Comparison of \prettyref{eq:fv3} and \prettyref{eq:cavA6}
shows that $V_{\!\text{qA}}$ passes the \emph{Einstein Addition Test.}

\section*{Appendix IV: Detailed Derivations of Eq.(\ref{eq:defB4}) and Eq.(\ref{eq:mom2}).}

To derive \prettyref{eq:defB4}, we have \prettyref{eq:defA1}, \prettyref{eq:defB2},
\prettyref{eq:detB3}, and $\left(\br V\cdot\br E\right)=\left(\br V\cdot\br B\right)=0$.
Using \prettyref{eq:detB3},
\[
\br S'=c\left(\br E'\times\br B'\right)\circeq c\gamma_{\!\text{B}}^{2}\left\{ \left(\br E\times\br B\right)+\br f+\br g\right\} 
\]
where, omitting zero terms,
\begin{align*}
\br f & =-\br E\times\left(\dfrac{\br V_{\!\text{B}}}{c}\times\br E\right)+\left(\dfrac{\br V_{\!\text{B}}}{c}\times\br B\right)\times\br B\\
 & =-\left(E^{2}+B^{2}\right)\dfrac{\br V_{\!\text{B}}}{c}=-\lambda\left(E^{2}+B^{2}\right)\dfrac{\br V_{\!\text{A}}}{c}\\
 & =-\lambda\left(E^{2}+B^{2}\right)\dfrac{2\left(\br E\times\br B\right)}{\left(E^{2}+B^{2}\right)}=-2\lambda\left(\br E\times\br B\right)
\end{align*}
and, again omitting zero terms,
\begin{align*}
\br g & =-\left(\frac{\br V_{\!\text{B}}}{c}\times\br B\right)\times\left(\frac{\br V_{\!\text{B}}}{c}\times\br E\right)\\
 & =-\dfrac{\br V_{\!\text{B}}}{c}\left\{ \left(\dfrac{\br V_{\!\text{B}}}{c}\times\br B\right)\cdot\br E\right\} =\dfrac{\br V_{\!\text{B}}}{c}\left\{ \dfrac{\br V_{\!\text{B}}}{c}\cdot\left(\br E\times\br B\right)\right\} \\
 & =\lambda^{2}\left\{ \dfrac{2\left(\br E\times\br B\right)}{\left(E^{2}+B^{2}\right)}\right\} \left\{ \dfrac{\br V_{\!\text{A}}}{c}\cdot\left(\dfrac{E^{2}+B^{2}}{2}\right)\dfrac{\br V_{\!\text{A}}}{c}\right\} \\
 & =\lambda^{2}\left(\dfrac{\br V_{\!\text{A}}}{c}\cdot\dfrac{\br V_{\!\text{A}}}{c}\right)\left(\br E\times\br B\right)=\lambda^{2}\left(\dfrac{V_{\!\text{A}}}{c}\right)^{2}\left(\br E\times\br B\right)
\end{align*}
Collect terms and factor out $\left(\br E\times\br B\right)$ to get
\[
\br S'=c\left(\br E'\times\br B'\right)\circeq\gamma_{\!\text{B}}^{2}c\left(\br E\times\br B\right)\left\{ \left(\dfrac{V_{\!\text{A}}}{c}\right)^{2}\lambda^{2}-2\lambda+1\right\} 
\]
 which is \prettyref{eq:defB4}.

To derive \prettyref{eq:mom2} we have \prettyref{eq:defA1}, \prettyref{eq:detB3},
and $\left(\br V\cdot\br E\right)=\left(\br V\cdot\br B\right)=0$.
Using \prettyref{eq:detB3}, and property (b) of footnote \ref{fn:cireqSymbol},
\begin{align*}
E'^{2} & =\gamma_{\!\text{B}}^{2}\left(\br E+\dfrac{\br V_{\!\text{B}}}{c}\times\br B\right)\cdot\left(\br E+\dfrac{\br V_{\!\text{B}}}{c}\times\br B\right)\\
 & =\gamma_{\!\text{B}}^{2}\left\{ E^{2}+2\br E\cdot\left(\dfrac{\br V_{\!\text{B}}}{c}\times\br B\right)+\left(\dfrac{\br V_{\!\text{B}}}{c}\times\br B\right)\cdot\left(\dfrac{\br V_{\!\text{B}}}{c}\times\br B\right)\right\} 
\end{align*}
Omitting zero terms,
\[
2\br E\cdot\left(\dfrac{\br V_{\!\text{B}}}{c}\times\br B\right)=-2\dfrac{\br V_{\!\text{B}}}{c}\cdot\left(\br E\times\br B\right)\quad\quad\text{and}
\]
\[
\left(\dfrac{\br V_{\!\text{B}}}{c}\times\br B\right)\cdot\left(\dfrac{\br V_{\!\text{B}}}{c}\times\br B\right)=\left(\dfrac{\br V_{\!\text{B}}}{c}\right)\cdot\left\{ \br B\times\left(\dfrac{\br V_{\!\text{B}}}{c}\times\br B\right)\right\} =\left(\dfrac{V_{\!\text{B}}}{c}\right)^{2}B^{2}
\]
Thus 
\[
E'^{2}=\gamma_{\!\text{B}}^{2}\left\{ E^{2}-2\dfrac{\br V_{\!\text{B}}}{c}\cdot\left(\br E\times\br B\right)+\left(\dfrac{V_{\!\text{B}}}{c}\right)^{2}B^{2}\right\} 
\]
Similarly,
\[
B'^{2}=\gamma_{\!\text{B}}^{2}\left\{ B^{2}-2\dfrac{\br V_{\!\text{B}}}{c}\cdot\left(\br E\times\br B\right)+\left(\dfrac{V_{\!\text{B}}}{c}\right)^{2}E^{2}\right\} 
\]
Combining, and using $\left(\br E\times\br B\right)=\left\{ 2{\cal E}/\left[1+\left(V_{\!\text{B}}/c\right)^{2}\right]\right\} \dfrac{\br V_{\!\text{B}}}{c}$
from eqs.(\ref{eq:defA1}, \ref{eq:defB2}, and \ref{eq:defB10}),
where ${\cal E}=\left(E^{2}+B^{2}\right)/2$, gives
\begin{align*}
{\cal E}' & =\dfrac{1}{2}\left(E'^{2}+B'^{2}\right)\\
 & =\gamma_{\!\text{B}}^{2}\left\{ \left[1+\left(V_{\!\text{B}}/c\right)^{2}\right]\dfrac{E^{2}+B^{2}}{2}-2\dfrac{\br V_{\!\text{B}}}{c}\cdot\left(\br E\times\br B\right)\right\} \\
 & =\dfrac{\gamma_{\!\text{B}}^{2}{\cal E}}{1+\left(V_{\!\text{B}}/c\right)^{2}}\left\{ \left[1+\left(V_{\!\text{B}}/c\right)^{2}\right]^{2}-4\left(V_{\!\text{B}}/c\right)^{2}\right\} \\
 & =\dfrac{\gamma_{\!\text{B}}^{2}{\cal E}}{1+\left(V_{\!\text{B}}/c\right)^{2}}\left[1-\left(V_{\!\text{B}}/c\right)^{2}\right]^{2}={\cal E}\,\dfrac{1-\left(V_{\!\text{B}}/c\right)^{2}}{1+\left(V_{\!\text{B}}/c\right)^{2}}
\end{align*}
which is \prettyref{eq:mom2}.

\section*{Appendix V: Detail of the Comoving System}

The comoving system is defined by $\br S'=c\left(\br E'\times\br B'\right)=0$.
Thus $\left|\br E'\times\br B'\right|=E'B'\sin\theta'=0$ where $\theta'$
is the angle between $\br E'$ and $\br B'$ in the comoving system. 

From eqs.(7.62 and 7.63) of Rindler \cite{rindler}, $({E'}^{2}-{B'}^{2})=(E^{2}-B^{2})$
and $(\br E'\cdot\br B')=(\br E\cdot\br B)$. It follows that: 

(a) An event with $(\br E\cdot\br B)\neq0$ has $E'B'\neq0$ and therefore
$\br E'$ and $\br B'$ must be either parallel or anti-parallel,
$\theta'=0$ or $\theta'=\pi$ at this event; 

(b) An event with $0=(\br E\cdot\br B)=(\br E'\cdot\br B')=E'B'\cos\theta'$
cannot have $E'B'\neq0$ in the comoving system because that would
require both $\cos\theta'=0$ and $\sin\theta'=0$. Thus $E'B'=0$
and one of $E'$ and $B'$ must be zero. If $E>B$ then $E'>B'$ and
hence $B'=0.$ If $E<B$ then $E'<B'$ and hence $E'=0$; 

(c) If both $0=(\br E\cdot\br B)$ and $E=B\neq0$ at an event, then
both $E'B'=0$ and $E'=B'$, and therefore $E'=B'=0$ and the fields
and energy density ${\cal E}'$ in the comoving system are zero. But
\prettyref{eq:defA1} and \prettyref{eq:defB8} show that such an
event also has $\left(V_{\!\text{A}}/c\right)=1$ and hence $\left(V_{\!\text{B}}/c\right)=1$
which is an unphysical value for a Lorentz boost velocity. The case
$E=B\neq0$ and $0=\br E\cdot\br B$ therefore must be approached
as a limit.

\bibliographystyle{plain}

\end{document}